\documentclass[10pt]{article}

\usepackage{doublespace}
\usepackage{note}

\begin{document}
\title{More Is Not Necessarily Easier}
\author{Debbie W. Leung and John A. Smolin} 
\address{I.B.M. T.J. Watson Research Center, P.O. Box 218, Yorktown 
Heights, NY 10598. \\
{\rm email wcleung@watson.ibm.com, smolin@watson.ibm.com}}
\maketitle
\centerline{\today}

\begin{abstract} 

In this brief comment, we consider the exact, deterministic, and
nonasymptotic transformation of multiple copies of pure states under
LOCC.  It was conjectured in quant-ph/0103131 that, if $k$ copies of
$|\psi\>$ can be transformed to $k$ copies of $|\phi\>$, the same
holds for all $r \geq k$.
We present counterexamples to the above conjecture.  

\end{abstract}

\section{Background and notations}

We consider pure bipartite states shared by Alice and Bob.  Let $A$
and $B$ be the labels of their respective quantum systems, each with
$d$ dimensions.  
Suppose Alice and Bob want to transform a state $|\psi\>$ to another
state $|\phi\>$.  They each have complete classical descriptions of
the states, but they can only perform Local quantum Operations and
Classical Communication (LOCC)~\cite{locc}.
The transformation can be performed under many scenarios.  The final
state can be exactly or approximately the desired output, and the
transformation can work deterministically or with certain probability.
It is also possible to consider converting between an asymptotically 
large numbers of two different states. 

We consider here {\em only} the scenario of exact, deterministic, and
nonasymptotic transformation using LOCC.
If such transformation is possible, we write $|\psi\> \rightarrow |\phi\>$, 
otherwise, we write $|\psi\> \not\rightarrow |\phi\>$.
A necessary and sufficient condition for $|\psi\> \rightarrow |\phi\>$
was found by Nielsen~\cite{Nielsen99}, with beautiful connections to the
theory of majorization.  For a pure state $|\eta\>$, let
$\lambda_\eta$ denote the list of eigenvalues of the reduced density matrix
${\rm Tr}_A (|\eta\>\<\eta|)$ in descending order, and let
$\lambda_\eta^{(i)}$ be the $i$-th element in $\lambda_\eta$.  
Nielsen's criterion can be stated as
\be
	|\psi\> \rightarrow |\phi\> 
	\hspace*{5ex} \mbox{iff} \hspace*{5ex} 
	\forall_{m \leq d}  \hspace*{2ex}
	\sum_{i=1}^m \lambda_\psi^{(i)} \leq \sum_{i=1}^m \lambda_\phi^{(i)}
\ee
An intriguing consequence of the criterion is that, even when $|\psi\>
\not\rightarrow |\phi\>$, it may still be possible to transform the
state $|\psi\>$ to the state $|\phi\>$ as part of a larger
transformation.
For example, $|\psi\>|\alpha\>\rightarrow |\phi\>|\alpha\>$ may be
true for some state $|\alpha\>$, which functions as a 
catalyst~\cite{Jonathan99}.  
It is also possible for $|\psi\>^{\otimes 2} \rightarrow
|\phi\>^{\otimes 2}$~\cite{many} or even $|\psi\>^{\otimes 4}
\rightarrow |\phi\>^{\otimes 5}$~\cite{Beckmann99}.  

In Ref.~\cite{Bandyopadhyay01}, interesting properties about 
multiple-copy transformation between $|\psi\>^{\otimes k}$ and
$|\phi\>^{\otimes k}$ under various scenarios are investigated.
In the scenario of exact, deterministic, asymptotic transformation
considered in this paper, they have derived 
a necessary condition for $|\psi\>^{\otimes
k} \rightarrow |\phi\>^{\otimes k}$ for some $k$,
$\lambda_\psi^{(1)} \leq \lambda_\phi^{(1)}$ and $\lambda_\psi^{(d)}
\geq \lambda_\phi^{(d)}$.
They have also made a very plausible conjecture, 
\be 
    |\psi\>^{\otimes k} \rightarrow |\phi\>^{\otimes k}
    \hspace*{2ex} \Longrightarrow \hspace*{2ex}  \forall_{r \geq k}
    \hspace*{2ex}
    |\psi\>^{\otimes r} \rightarrow |\psi\>^{\otimes r}.
\ee
The conjecture is obviously true when $k$ divides $r$.
However, we find counterexamples for $k=2$, $r=3$, and $d \geq 5$.  
One example for $d = 5$ is given by 
\bea
	\lambda_\psi = \{0.493 \,, 0.284 \,, 0.158 \,,  0.035 \,, 0.030 \}
\nonumber
\\	\lambda_\phi = \{0.598 \,, 0.145 \,, 0.129 \,,  0.125 \,, 0.003 \}
\nonumber
\eea
and a simple example for $d = 6$ is given by 
\bea
	\lambda_\psi = \{0.24\,, 0.22\,, 0.22 \,, 0.19 \,, 0.10 \,, 0.03 \}
\nonumber
\\	\lambda_\phi = \{0.27\,, 0.25\,, 0.16 \,, 0.16 \,, 0.15 \,, 0.01 \}
\nonumber
\eea
The conjecture fails because of the exact and nonasymptotic nature of
the transformation.  

We remark that, for fixed $k$, a counterexample in $d_0$ dimensions
can always be turned into one in $d \geq d_0$ dimensions.  The reason
is that the criteria for a counterexample involve only continuous
functions of $\lambda_\psi$ and $\lambda_\phi$.
Therefore, the above counterexamples suffice to disprove the conjecture 
for $k=2$ and $\forall d \geq 5$.  
In the following, we describe numerical results that provide not only
counterexamples but also statistics of violations of the conjecture.

\section{Counterexamples}

We performed a numerical search for $d = 4, 5, \ldots, 9$ and $k=2$.  
The $d = 3$ case is uninteresting, as 
$|\psi\> \not\rightarrow |\phi\>$ implies 
$\forall_k~|\psi\>^{\otimes k} \not\rightarrow |\phi\>^{\otimes k}$.
For each $d$, a large number of random pairs $\{ \lambda_\psi,
\lambda_\phi \}$ are sampled, and without loss of generality,
$\lambda_\psi^{(1)} \leq \lambda_\phi^{(1)}$.  We search for $\{
\lambda_\psi, \lambda_\phi \}$ corresponding to $|\psi\>, |\phi\>$
satisfying 
\\[1.5ex]
\hspace*{5ex} {\bf I~} $|\psi\> \not\rightarrow |\phi\>$ and 
$|\psi\>^{\otimes 2} \rightarrow |\phi\>^{\otimes 2}$  
\\[1.3ex] 
\hspace*{5ex} {\bf II} 
$|\psi\>^{\otimes 3} \not\rightarrow |\phi\>^{\otimes 3}$ 
\\[1.5ex]
The search parameters and the results can be summarized: 
\begin{table}[h]
$
\begin{array}{l| l| l| l| l| l| l} 
& d = 4 & d = 5 & d = 6 & d = 7 & d = 8 & d = 9 \\[0.3ex]
\hline
\rule{0pt}{2.4ex} 
\begin{array}{l} \mbox{Sample size} \end{array} & 
4 \times 10^8 & 2 \times 10^8 & 10^7 & 10^7 & 10^7 & 10^7 \\
\hline
\begin{array}{l} \mbox{Fraction of sample} 
              	\\ \mbox{satisfying {\bf I}} \end{array} & 
8.31 \times 10^{-3} & 1.56 \times 10^{-2} & 
2.16 \times 10^{-2} & 2.63 \times 10^{-2} & 
3.01 \times 10^{-2} & 3.31 \times 10^{-2} \\
\hline
\begin{array}{l} \mbox{Fraction of sample}
		\\ \mbox{satisfying {\bf II}} 
		\mbox{ given {\bf I}} \end{array} & 
0 & 3.21 \times 10^{-7} & 
3.94 \times 10^{-4} & 3.12 \times 10^{-4} & 
1.60 \times 10^{-4} & 1.63 \times 10^{-4} \\
\end{array}
$ 
\end{table}

The fractions of sample satisfying {\bf I} have negligible statistical
uncertainties, while the fractions satisfying {\bf II} conditioned on {\bf
I} have about 11-14\% uncertainties\footnote{Except for $d=4$ and $5$ for
which few counterexamples are found.}.  For each $d \geq
6$, tens of counterexamples for the conjecture are found.  
For $d = 5$, exactly one counterexample is found in $2 \times 10^{8}$
random samples.  For $d=4$ no counterexample is found in $4 \times
10^{8}$ random samples, so that the conjecture may still be true for
this value of $d$.

\section{Discussion}

The conjecture investigated in this paper fails because of the exact, 
deterministic, and nonasymptotic nature of the transformation.
The conclusion drawn from our results, along with 
Refs.~\cite{many,Beckmann99,Bandyopadhyay01}, is that 
even if one knows for all $k < r$, whether
$|\psi\>^{\otimes k} \rightarrow |\phi\>^{\otimes k}$, one still
cannot infer whether $|\psi\>^{\otimes r} \rightarrow |\phi\>^{\otimes
r}$ without additional information on the states.  
This is in sharp contrast with the much simpler conditions for
asymptotic manipulations.  
In particular, asymptotically $|\psi\>^{\otimes m}$ and 
$|\phi\>^{\otimes n}$ are interconvertible with high fidelity and probability  
if $m H(\lambda_\psi) = n H(\lambda_\phi)$ 
where $H$ is the entropy function. 

Nielsen's criterion on transformation of pure bipartite states 
has led to many important results in the various possible scenarios, 
some of which are counterintuitive.  
The results in the exact, deterministic, nonasymptotic scenario are
mathematically interesting, yet their very oddity suggests that the
asymptotic or probabilistic scenarios are perhaps more physically
relevant.

\section{Acknowledgments}

We thank Charles~H.~Bennett for interesting discussions and the IBM Watson 
cafeteria for enlightening coffee.

\end{document}